\documentclass[prb,a4paper,twocolumn,showpacs,floatfix]{revtex4-1}
\usepackage[utf8]{inputenc}
\usepackage[T1]{fontenc}
\usepackage{lmodern}
\usepackage{amssymb}
\usepackage{amsmath}
\usepackage{amsbsy}
\usepackage{bm}
\usepackage{esint}
\usepackage{graphicx}
\graphicspath{ {figs/} }
\usepackage{color}
\usepackage{url}
\usepackage{siunitx}

    \newcommand{\I}{\mathrm{i}}
    \newcommand{\E}{\mathrm{e}}
    
    \renewcommand{\d}[1]{\mathrm{d}#1\,}

\begin{document}
\title{Skyrmion collapse}
\author{Alberto D.\ Verga}\email{Alberto.Verga@univ-amu.fr}
\affiliation{Aix-Marseille Université, IM2NP, Campus St Jérôme, service 142, 13387 Marseille, France}
\date{\today}
\begin{abstract}
We investigate the topological change in a Belavin-Polyakov skyrmion under the action of a spin-polarized current. The dynamics is described by the Schrödinger equation for the electrons carrying the current coupled to the Landau-Lifshitz equation for the evolution of the magnetic texture in a square lattice. We show that the addition of an exchange dissipation term, tends to smooth the transition from the skyrmion state to the ferromagnetic state. We demonstrate that this topological change, in the continuum dissipationless limit, can be described as a self-similar finite-time singularity by which the skyrmion core collapses.
\end{abstract}
\maketitle

\section{Introduction}

The equilibrium magnetization field in ferromagnetic nanodots and in chiral magnets, often possesses a nontrivial topology. Magnetic vortices in permalloy\cite{Van-Waeyenberge-2006fk} and skyrmion lattices in transition metal compounds\cite{Muhlbauer-2009vn} were experimentally observed. These configurations are interesting because a change between states having different topologies (switching of the vortex core, motions and annihilation of skyrmions), can be used as basic states in non-volatile memories and spintronic devices.\cite{Romming-2013lq} A generic physical mechanism that can trigger these topological changes, without magnetic fields, is the spin-transfer torque.\cite{Ralph-2008ly} It consists of the interaction between itinerant spins, produced by a spin-polarized current, and magnetic moments (fixed spins) of the magnetic material, when the underlying magnetization is non-uniform.

From a more general point of view, physical systems involving defects or topological singularities are among the more rich and interesting to investigate; the difficulty to consistently deal with these singularities resides in the coupling of a large range of scales relevant to describe their structure and dynamics. For instance, in the domain of magnetic textures in nanoscale systems, the switching of a magnetic vortex entails the nucleation of a Bloch point, which is a magnetization field singularity.\cite{Hertel-2006ib,Elias-2011uq} The characteristic length, time and energy scales involved in the formation of the Bloch point range from the crystal lattice, where the dominant effect is the neighboring spins exchange interaction, to the intermediate micromagnetic structures, and up to the geometry of the ferromagnet through the dipolar magnetic field.\cite{Andreas-2014qf} A similar difficulty appears in the transition from a skyrmion state towards a ferromagnetic state in chiral magnets, driven by a localized electric current pulse, where spin-orbit together with exchange couplings are present.\cite{Romming-2013lq,Sampaio-2013xy} 

The dynamics of magnetic vortices is usually studied in the micromagnetic approximation, based on the Landau-Lifshitz equation.\cite{Landau-1935fk} Within this framework, to account for the spin-transfer torque one has to extend the Landau-Lifshitz equation with terms proportional to the current and the magnetization gradients.\cite{Zhang-2004ve} However, this approach neglects strong non-adiabatic effects, such as the generation of current inhomogeneities due to the scattering of electrons on the magnetization gradients. In order to investigate topological changes in ferromagnets, we recently proposed a self-consistent model, where electrons obey quantum dynamics (see Ref.~\onlinecite{Elias-2014fk} and references therein).  

In this paper we develop this model further, focusing on the skyrmion-ferromagnetic transition induced by a spin polarized current. We are in particular interested in the relationship between topological change and dynamics. In analogy with the collapse of Langmuir solitons\cite{Zakharov-1972ty} we ask whether a finite time singularity would arise in the evolution of a driven localized magnetic structure. The basic magnetic texture we consider, is a stabilized version of the Belavin-Polyakov\cite{Belavin-1975xw} skyrmion, appropriated to a system defined on a periodic lattice. After a presentation of the basic equations coupling the Schrödinger equation for the itinerant spins (electrons) to the Landau-Lifshitz equation for the ferromagnetic spins (fixed classical magnetic moments), defined on a square lattice, we present a qualitative model to show the basic mechanism of the transition. The transition from the skyrmion state to the ferromagnetic state implies a change in topology, and thus a violation of the topological charge conservation.  We include an exchange dissipation term,\cite{Baryakhtar-1984fk} that in principle can smooth the transition. We performed a series of numerical computations to study the phenomenology of the transition and to identify the mechanism of the topological change, and their dependence on the dissipation. The main result of this paper, is the description of the topological change as a finite-time singularity in the dynamics of the skyrmion (in the continuum limit). A self-similar solution of the Landau-Lifshitz equation driven by the spin-torque term is found and compared to the texture observed in simulations.

\section{Model}

The motion of an electron in a lattice of step \(a\) and size \(L^2\) is given by the Heisenberg equation for the two components (for the spin-up and spin-down) annihilation operator \(c_i\) at site \(i\),
\begin{equation}\label{e:c}
  \I \hbar \, \dot{c}_i(t) = [c_i(t),H_e(t,S_i)]\,,
\end{equation}
where \(H_e\) is the time dependent hamiltonian,
\begin{multline}
  \label{e:He}
  H_e =  -\epsilon\sum_{\langle i,j\rangle} 
    \E^{\I \phi_{i,j}(t)}c_i^\dag c_j
    -J_s\sum_i \bm S_i \cdot (c_i^\dag \bm \sigma c_i)\\
    -\bm B_p \cdot \sum_i c_i^\dag \bm \sigma c_i
\end{multline}
where the fixed \(\bm S_i\) and itinerant spins \(c_i^\dag \bm \sigma c_i\) (\(\bm \sigma\) is the vector of Pauli matrices) are coupled by the exchange constant \(J_s\); \(\epsilon\) is the energy to jump from site \(i=(x_i, y_i)/a=\bm x_i/a\) to its neighbor \(j\). The system is subject to a constant electric field \(E\) in the \(x\)-direction, responsible for the phase factor appearing in the kinetic energy term, 
\[
  \phi_{i,j}(t)=(\bm x_i- \bm x_j)\cdot\hat{\bm x} \frac{eEt}{\hbar}
\] 
(with \(-e\) as the electron charge, and \(\hat{\bm x}\) as the unit vector in the \(x\)-direction). The last term in (\ref{e:He}) contains the current polarization effective magnetic field \(\bm B_p\) in energy units.

The magnetic texture follows the dynamics given by the Landau-Lifshitz equation,
\begin{equation}
  \label{e:LL}
  \hbar \frac{\partial}{\partial t}\bm S_i=
    \bm S_i\times(\bm f_i - \alpha \bm S_i \times \bm f_i + J_s \bm s_i) 
         - \bm d\,,
\end{equation}
where the effective field,
\begin{equation}
  \label{e:fi}
  \bm f_i = -\frac{\delta H_S}{\delta \bm S_i}\,,
\end{equation}
is derived from the coarse-grained free energy,
\begin{equation}
  \label{e:HS}
  H_S = \frac{J}{2} \sum_i (\nabla \bm S_i)^2
\end{equation}
\(J\) is the Heisenberg exchange constant, and \(\alpha\) is a damping constant. The last term \(\bm d\), includes a dissipation mechanism to be specified below. This equation is coupled to the equation for the electrons through the torque in \(J_s\), where the electron spin is computed by the formula \(\bm s_i(t)= \langle c^{\dag}_i(t) \bm \sigma c_i(t) \rangle\) where the bracket is for the quantum mean value. We use periodic boundary conditions in order to have a well defined topology. Note that \(\nabla\) is a difference operator acting on the lattice sites. In practice, the difference operators are computed in Fourier space and transformed to the lattice space. In units such that \(\epsilon=a=\hbar=e=1\), typical parameters are as follows: \(J_s=1\), \(J=0.4\), \(\alpha=B_p=n_e=0.1\), \(E=10^{-3}\) and \(L=128\). With these parameters, the order of magnitude of physical units are as follows: length \(a=\SI{1}{\nano \metre}\), time \(t_0=\SI{1}{\femto \second}\) (the magnetization characteristic time is \(\sim 10\,t_0\)), energy \(\epsilon=\SI{1}{eV}\), electric field \(E_0=\SI{1e9}{\volt \per \metre}\).

Without the dissipation term, the system (\ref{e:c}-\ref{e:HS}) conserves the magnetization modulus \(|\bm S|=1\), and the topological charge \(Q=Q(t)\),
\begin{equation}
  \label{e:Q}
  Q = \int \frac{\d{\bm x}}{4\pi} q(\bm x, t)\,,\quad
  q=\bm S\cdot\partial_x\bm S\times \partial_y \bm S\,,
\end{equation}
where the integration is over the lattice, and \(q\) is the topological charge density (note that the change \(S_z \rightarrow -S_z\), change the sign of \(Q\)). 

Topological changes, such as the transition between a skyrmion state and a ferromagnetic state, or the switch of vortex polarity, need, in the continuum limit, a transition state where the core magnetization vanishes at some point; this is impossible in the framework of the Landau-Lifshitz equation, which strictly conserves the norm of \(\bm S\). For instance, in the intermediate state of a three dimensional vortex switching, a Bloch point is nucleated.\cite{Hertel-2006ib} The inclusion of a dissipation mechanism relevant in the low temperature and strong magnetization gradients limit, is essential to allow a regularization of the dynamics. A simple phenomenological theory, allows for deriving a general form of an exchange driven dissipation term.\cite{Baryakhtar-1984fk} The rate of dissipation of the total magnetic energy, taken as a functional of the magnetic moments distribution \(H_S[\bm S]\), is as follows, 
\[
  \hbar\dot H_S = -\hbar \int \d{V} \bm f \cdot \dot{\bm S} =  
  \int \d{V} \bm f \cdot \bm d \,,
\]
where we used (\ref{e:fi}); the integration is on the system's volume (in units of the lattice cell size). The dissipation vector can be expanded, as in the linear response theory, in the gradients of the effective field \(\bm f\). In the simplest isotropic case, and taking into account that the corresponding term must appear as a divergence in the energy balance equation, the most general form is:
\begin{equation}
  \label{e:dis}
  \bm d = \beta \nabla^2 \bm f_i\,,
\end{equation}
with \(\beta\) as a positive nondimensional constant. This choice gives \(\hbar \dot H_S = - \beta \int \d{V} (\nabla \bm f)^2 < 0\). Dissipation effects not depending on gradients are already included in the terms proportional to \(\alpha\) in (\ref{e:LL}). The vector (\ref{e:dis}) describes the energy dissipation related to exchange interactions, and thus involving neighboring magnetic moments: \(\bm d \sim \beta J \nabla^4 \bm S\), is analogous to a curvature dissipation term. It breaks the conservation of the magnetization norm.

\section{Phenomenology of the transition}

%
\begin{figure*}
  \centering
  \includegraphics[width=0.22\textwidth]{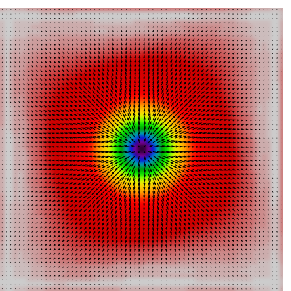}\hfill%
  \includegraphics[width=0.74\textwidth]{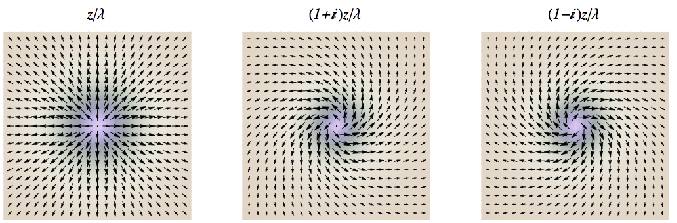}
  \caption{\label{f:skyr} (color online). Initial magnetization used in numerical computations in the form of a \(Q=-1\) skyrmion (left). Arrows give the magnetization in the plane, and color, gives its \(z\)-component: at the center \(\bm S=(0,0,-1)\). The three right panels show skyrmions with topological charge \(Q=1\) and size \(\lambda\) defined by the stereographic projection on the complex plane \(w=w(z)\), for (left) \(w=z/\lambda\), (center) \(w=(1+\I)z/\lambda\), and (right) \(w=(1-\I)z/\lambda\). The opposite charge is obtained by the transformation \(w \rightarrow 1/\bar w\), which changes \(S_z \rightarrow -S_z\). A polarized current induces a precession motion that changes the arrows orientation around the center in time. The lattice side is \(L=128a\) and \(\lambda=20a\).}
\end{figure*}

We performed numerical computations based on the integration of the coupled equations (\ref{e:c}) and (\ref{e:LL}) to investigate the transition between an initial nontrivial topological state and a final uniform magnetization state, driven by an electron current initially uniformly polarized. We use a full spectral model on the lattice to compute the spatial spin texture of electrons and classical spins. Time stepping of the Schrödinger equation for the electron quantum state \(|\psi(i,t)\rangle = c_i^\dag |0\rangle\), where \(|0\rangle\) is the electron initial state, use a splitting method that exactly conserves the norm of the quantum state.

The initial distribution of the magnetization is computed using a relaxation method to find a local equilibrium solution of the Landau-Lifshitz equation. We start with a Belavin-Polyakov skyrmion of topological charge \(Q=-1\):
\begin{equation}
  \label{e:bps}
  S_x = \frac{2\lambda r \cos\phi}{\lambda^2+r^2},\;
  S_y = \frac{2\lambda r \sin\phi}{\lambda^2+r^2},\;
  S_z = -\frac{\lambda^2 - r^2}{\lambda^2 + r^2}\,,
\end{equation}
of size \(\lambda \ll L\), with the central core pointing in the \(z\)-down direction, and with magnetization in the up direction at infinity (Fig.~\ref{f:skyr} left panel). This continuous and extending to infinity magnetization field, relaxes towards a periodic distribution defined on the square lattice. However, at scales \(\ell\) intermediate between the lattice and the spatial period lengths, \(a \ll \ell \ll L\), it is similar to the continuous distribution, as seen in Fig.~\ref{f:skyr}. The parameter \(\lambda\) is chosen to satisfy this requirement. The skyrmion is driven by an electron current, which is taken to be initially uniform, polarized in the up direction, parallel to the fixed spins orientation far from the core region.

\begin{figure*}
  \includegraphics[width=0.235\textwidth]{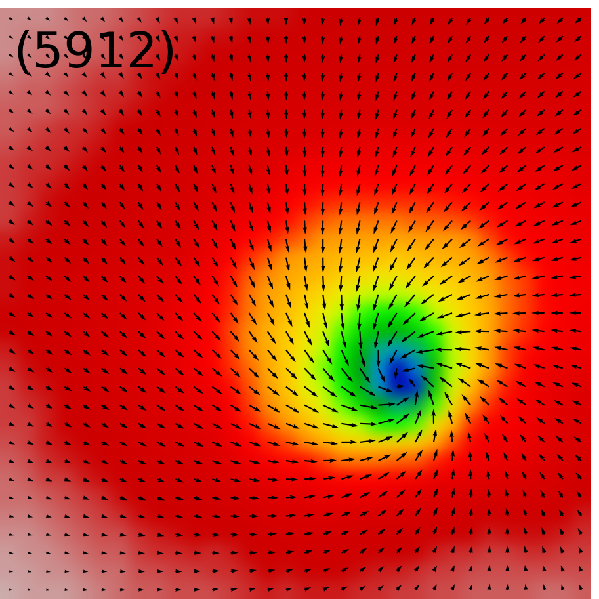}\hfill%
  \includegraphics[width=0.235\textwidth]{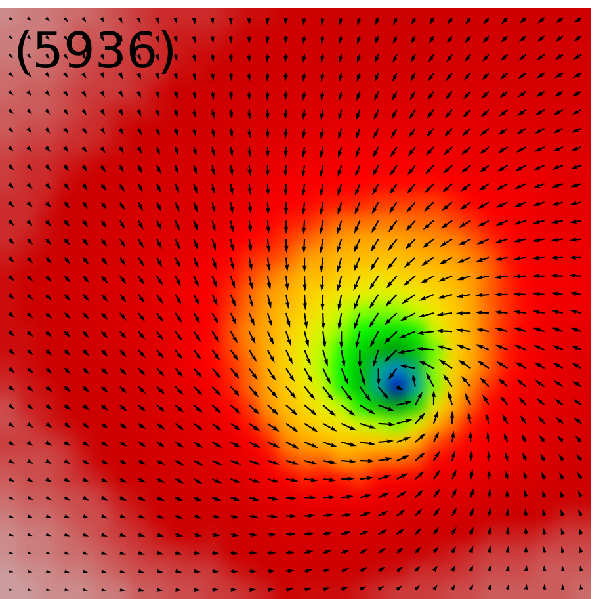}\hfill%
  \includegraphics[width=0.235\textwidth]{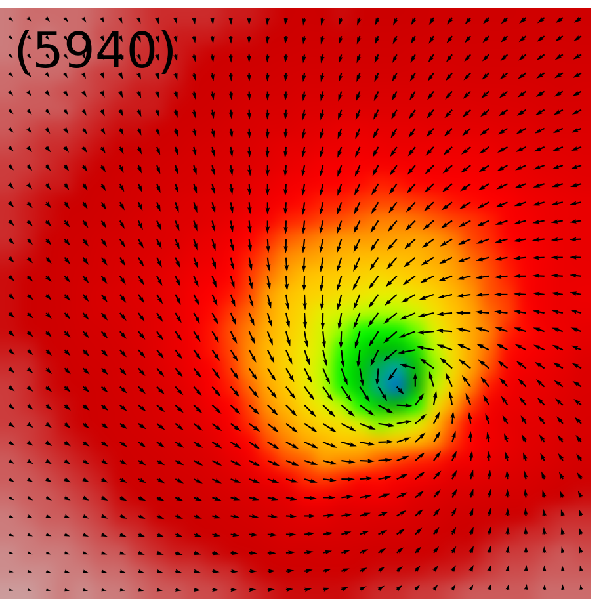}\hfill%
  \includegraphics[width=0.235\textwidth]{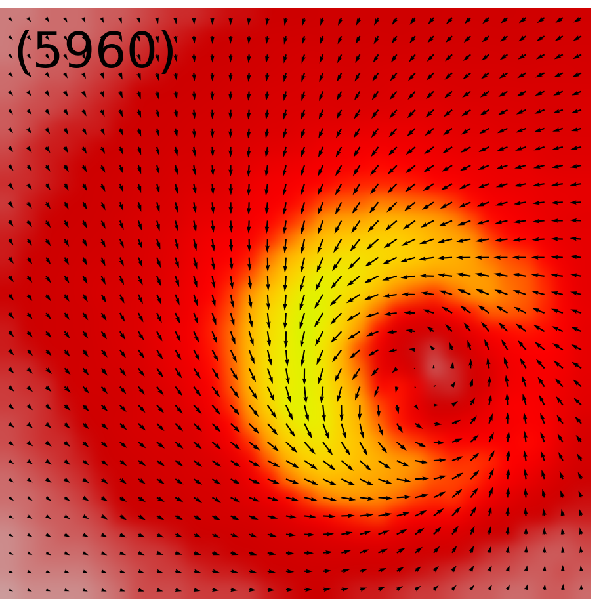}\hfill%
  \includegraphics[width=0.028\textwidth]{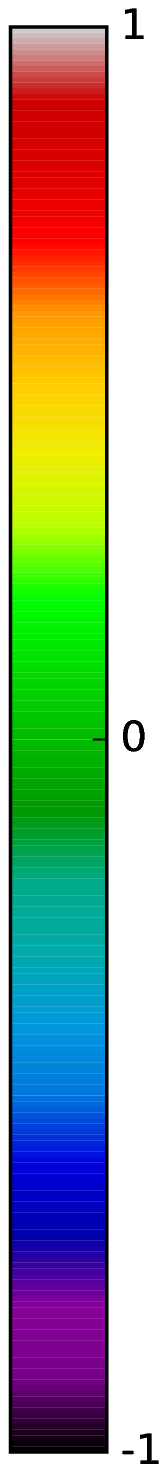}
  \caption{\label{f:tt} (color online). Magnetization field near the skyrmion core at the topological change (at time \(t \approx 5936\)). The core collapse (two left panels, \(t=5912,5936\,t_0\)) is followed by the emission spin waves (two right panels, \(t=5940,5960\,t_0\)). The magnetization is shown in full resolution, an arrow per lattice site, in a square of side 64$a$. Parameters: \(\beta=0.001\), \(E=10^{-3}E_0\). }
\end{figure*}

The numerical computations show an initial transient characterized by the deformation of the skyrmion core and the generation of spin waves if the electric field is strong enough. After this transient, a quasi-stationary state sets in. One observes a smooth precession motion of the fixed spins, and depending on the intensity of the current, a displacement of the skyrmion core. The topological charge is not affected by the transient or the subsequent regular dynamics of the magnetic texture. During the evolution of the skyrmion, its core size decreases. The reduction of the core dramatically accelerates at some given time, at which a transition to other state with different topology occurs. In Fig.~\ref{f:tt} we show snapshots of the skyrmion core region during the topological change.

%

\begin{figure*}
  \centering
  \includegraphics[width=0.32\textwidth]{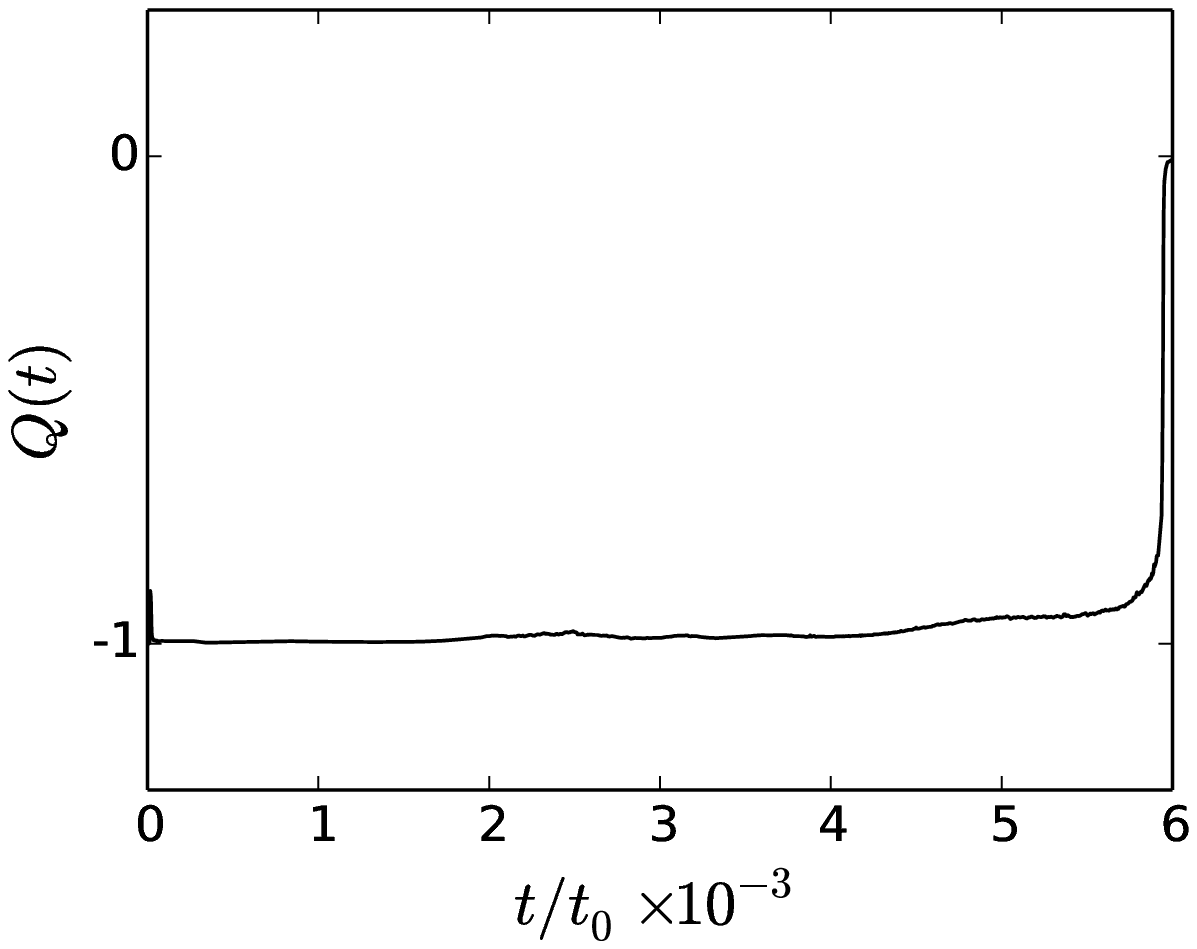}\hfill%
  \includegraphics[width=0.32\textwidth]{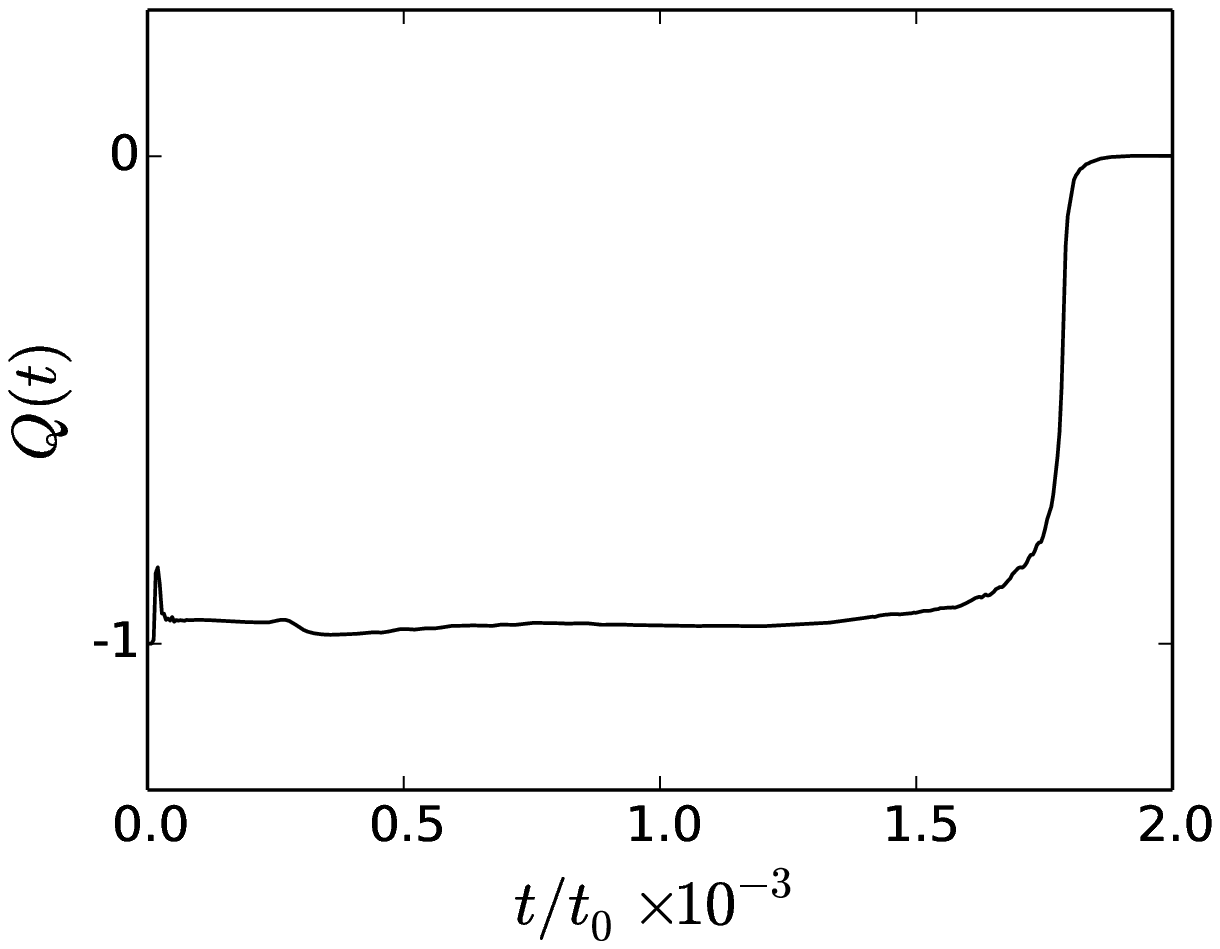}\hfill%
  \includegraphics[width=0.32\textwidth]{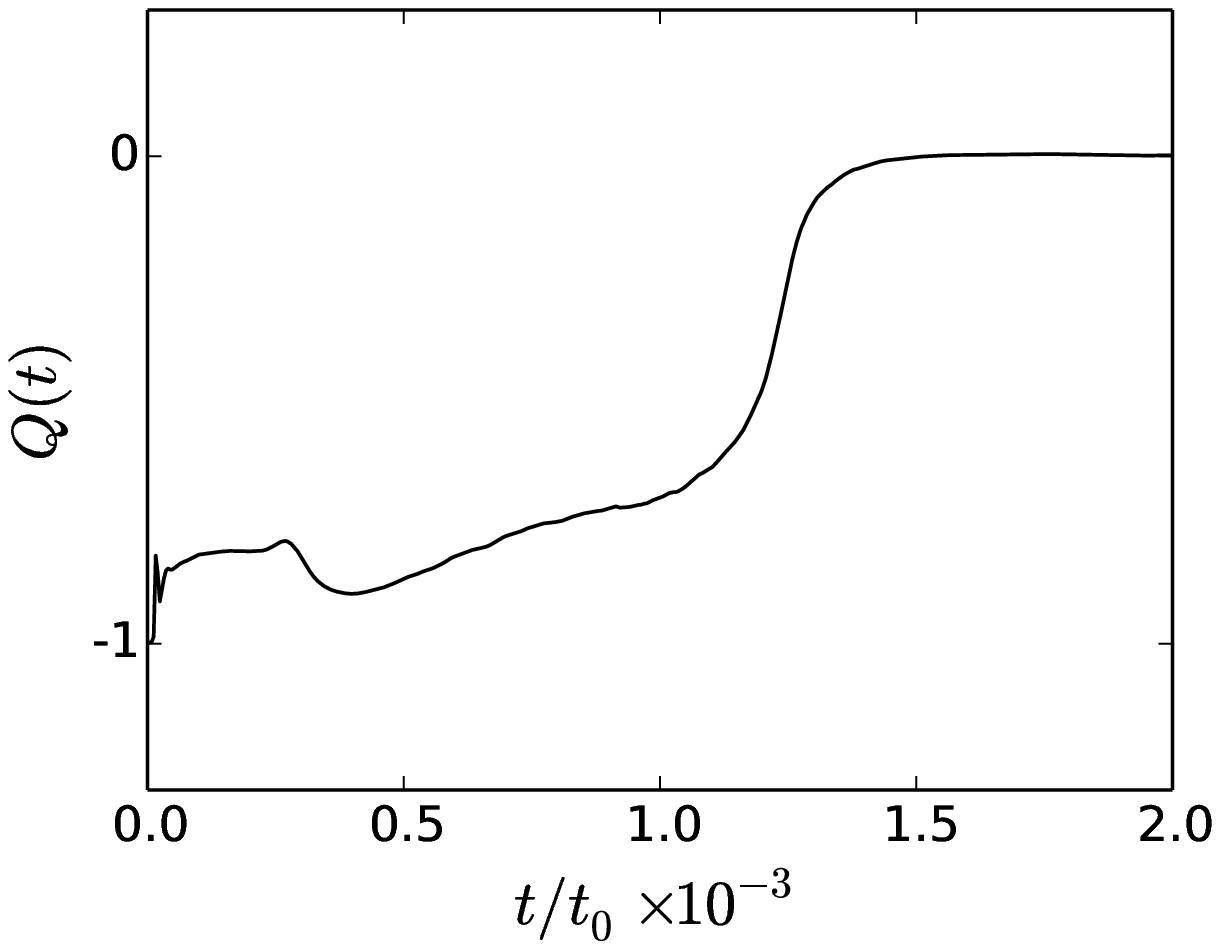}
  \caption{\label{f:q} Topological charge \(Q\) as a function of time (in adimensional units) for different values of the exchange dissipation parameter: (left) \(\beta = 0.001\), (center) \(\beta = 0.01\), and (right) \(\beta = 0.1\). The initial skyrmion charge is \(Q=-1\), and the current is polarized in the \(-z\) direction. The change from \(Q=-1\) to \(Q=0\) corresponds to a transition from the skyrmion to the ferromagnetic sate. Increasing the dissipation strength results in a decrease of the time necessary to reach the transition: \(t=\)\numlist{5936;1748;1236}, for \(\beta=\)\numlist{0.001;0.01;0.1}, respectively. The electric field is \(E=10^{-3}\).}
\end{figure*}

The scenario of the transition between the skyrmion and the ferromagnetic states, can be influenced by the dissipation. The first effect of the dissipation is to break the invariance of the topological charge, which is no longer restricted to take integer values. We observe in particular that the exchange dissipation term favorises the transition from the skyrmion state to the ferromagnetic state. Indeed, we plot in Fig.~\ref{f:q} the evolution of the topological charge as a function of the dissipation strength. In addition to the increasing instability of the skyrmion state with increasing dissipation, it is worth noting that the characteristic transition time becomes longer with dissipation. This is, as expected, a regularization effect. In the limit of vanishing dissipation the transition produces, ultimately, by the change of a single spin.\cite{Elias-2014fk}

\begin{figure*}
  \centering
  \includegraphics[width=0.32\textwidth]{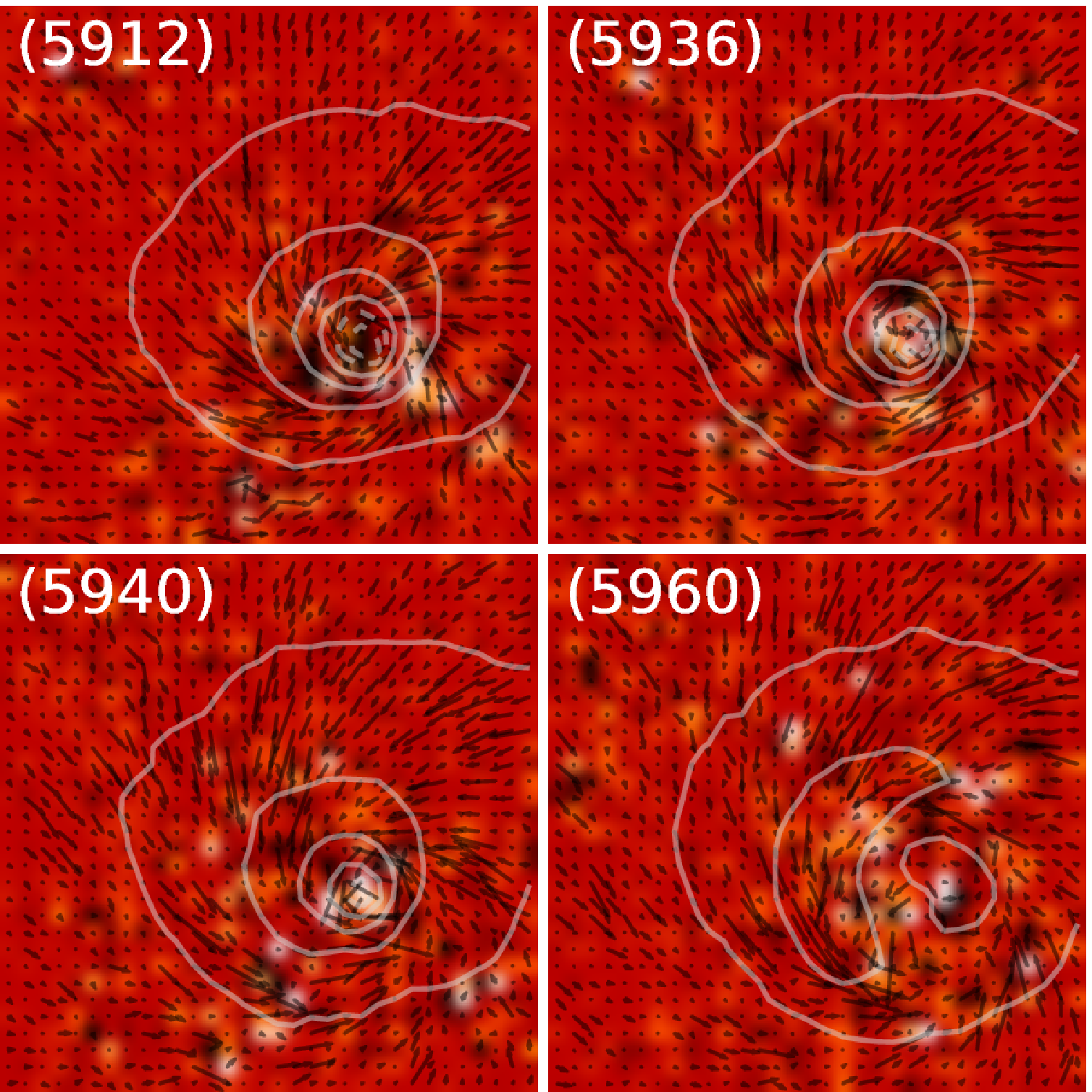}\hfill%
  \includegraphics[width=0.32\textwidth]{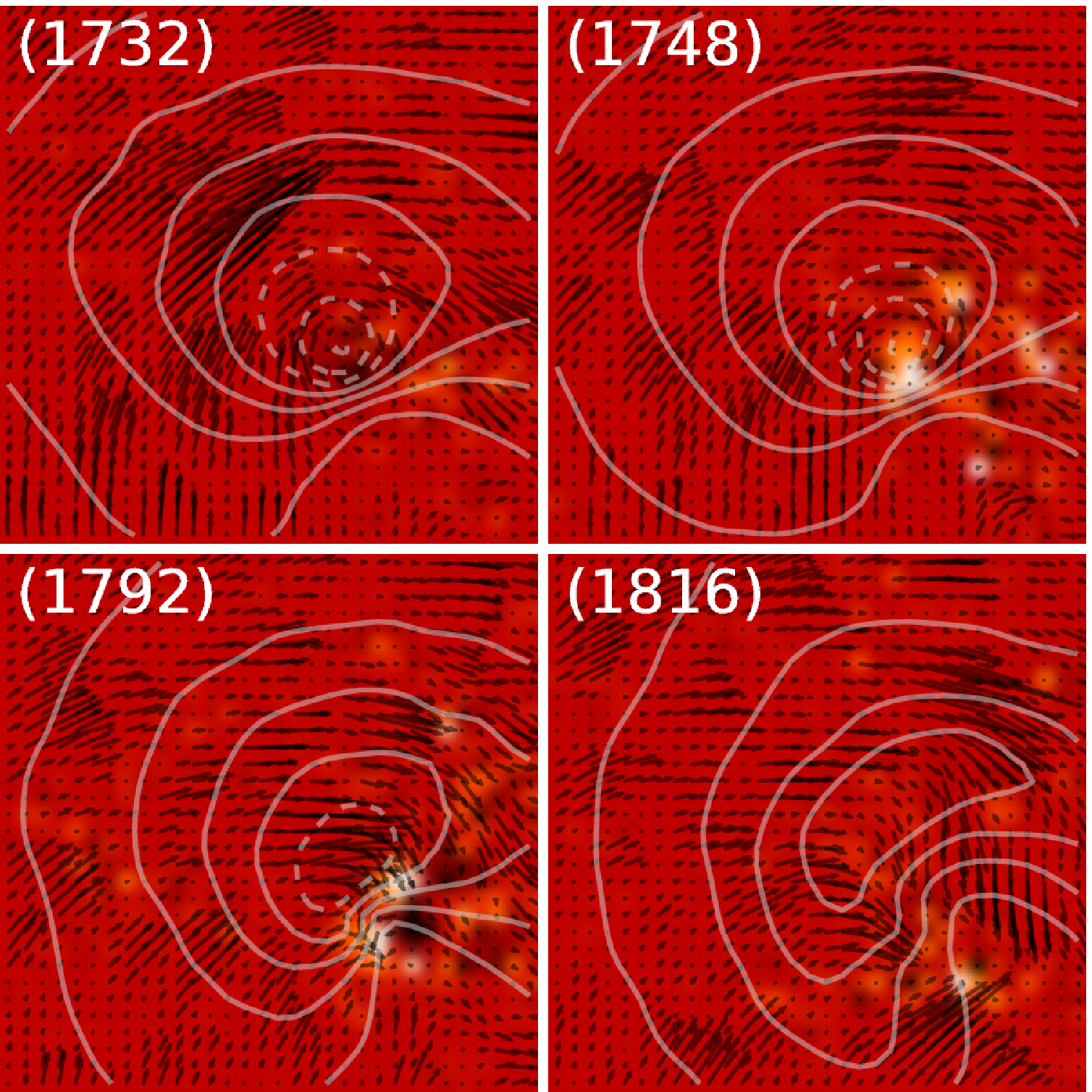}\hfill%
  \includegraphics[width=0.32\textwidth]{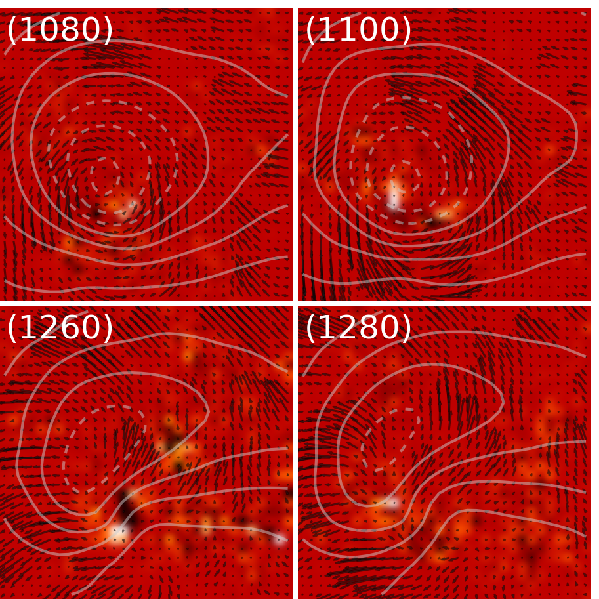}\\
  \includegraphics[width=0.32\textwidth]{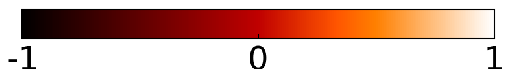}
  \caption{\label{f:b} Phenomenology of the topological transition. Contours of \(S_z\) (fixed spins), arrows of \((s_x,s_y)\) (itinerant spins), and color density of the topological \(b\)-field. The dissipation is (left) \(\beta = 0.001\), (center) \(\beta = 0.01\), and (right) \(\beta = 0.1\). The transition towards the ferromagnetic state is correlated with the appearance of intense \(b\)-field structures possessing a topological charge opposite to the one of the original skyrmion.}
\end{figure*}

A quantitative explanation, from the topological point of view, of the mechanism behind the transition is based on the behavior of the electron spins. We discuss in the following section, the dynamical point of view of the transition. As we can observe in Fig.~\ref{f:b}, although the electrons dynamics is almost stochastic (due to multiple scattering and interference effects), their spins organize near the skyrmion core. This can be verified by measuring the topological \(b\)-field, defined in a similar way as the density \(q\) of topological charge, but substituting \(\bm s\) to \(\bm S\):
\begin{equation}\label{e:b}
b=\bm n\cdot\partial_x\bm n 
	\times \partial_y \bm n,\quad \bm n=\bm s/|\bm s|\,;
\end{equation}
represented in Fig.~\ref{f:b} by the color density. We verify that the transition is associated with the nucleation of a well localized electron vortex having a topological charge density opposite to the one of the background skyrmion (the white spots that appear near the skyrmion core at times \(t= 5936,\, 1748,\, 1100\), for the three values of the dissipation, respectively). We remark, that despite the dissipation, the characteristic length scale of these structures is comparable with a few lattice steps, confirming their topological nature (as defects in the elastic field of a crystal). These structures, highly fluctuating in the low dissipation limit, tend to stabilize at high dissipation. In particular, the white spot of localized \(b\)-field, appearing at \(t\approx 1100\) in the case \(\beta=0.1\) of Fig.~\ref{f:b} (right panel), is always present at \(t=1260\), before fading out. The behavior of the electron spin field is similar in the intermediate dissipation case of \(\beta = 0.01\), but on a much shorter time scale. In the three cases we observe a strong spin field parallel to the plane in the neighborhood of the skyrmion center forming a vortex like structure. 

These observations naturally lead to a scenario in which the topological change is related to the interaction of the skyrmion core spins with this topological field. In such a case, the topological change can be accounted for as a balance equation: \(Q_{\text{final}} = Q_{\text{sky}} + Q_{\text{elec}} = -1 + 1 = 0\) the final charge results from the addition of the skyrmion charge and the electron spin structure charge. This supposes that the fixed spins are at the transition time, looked to the electron spins dynamics (the opposite of the adiabatic assumption). The interaction should result then in the annihilation of the skyrmion topological charge by the nucleation of an electron structure of opposite charge, accomplishing in this way, the transition towards a topologically different state. In what follows, we attempt to relate this topological change to the dynamics of the magnetization under the action of the electron itinerant spins.

%
%

\section{Skyrmion}

To study the dynamics of the skyrmion-ferromagnetic transition it is convenient to work in the dissipationless continuous limit, and to transform the magnetization field (which has only two independent components) using the stereographic projection:
\begin{equation}
  \label{e:stereo}
  S_x = \frac{w+\bar{w}}{1+|w|^2}\,,\;
  S_y = \frac{1}{\I} \frac{w-\bar{w}}{1+|w|^2}\,,\;
  S_z = \frac{1-|w|^2}{1+|w|^2}\,,
\end{equation}
with 
\begin{equation}
  \label{e:istereo}
  w = \frac{S_x+\I S_y}{1+S_z}\,.
\end{equation}
When \(S_z=1\), \(w\) goes to zero, and in the opposite pole, \(S_z=-1\), \(w\) goes to infinity. This projection maps the vector field over the unit sphere \(\bm S =\bm S(\bm x,t)\) to the field over the complex plane \(w = w(z,\bar{z},t)\) (where \(z=x+\I y\)). The Landau-Lifshitz equation (\ref{e:LL}) with \(\bm d=0\) and \(\alpha=0\), becomes,
\begin{multline}
  \label{e:wt}
  \I \partial_t w = -J\partial \bar{\partial} w + 
  \frac{2 J \bar{w}}{1+|w|^2} \partial w \bar{\partial} w \\
  - \tfrac{1}{2}s_+ +  s_z w + \tfrac{1}{2} s_- w^2\,,
\end{multline}
(only the exchange field in \(J\), is considered here) where we defined the complex derivative \(\partial=\partial/\partial z=\tfrac{1}{2}(\partial/\partial x- \I \partial/\partial y)\) and its complex conjugate \(\bar{\partial}\). The second line in (\ref{e:wt}) corresponds to the spin-transfer torque term, where \(s_\pm = s_x \pm \I s_y\). The exchange energy and the topological charge, in the stereographic representation, are given by,
\begin{align}
  E_{xc} &= 2J \int \d{z}\d{\bar z}
      \frac{|\partial w|^2 + |\bar\partial w|^2}{(1+|w|^2)^2}\,,
      \nonumber \\
  Q &= \frac{1}{2\pi} \int \d{z}\d{\bar z}
      \frac{|\partial w|^2 - |\bar \partial w|^2}{(1+|w|^2)^2}\,.
  \label{e:excq}
\end{align}
Equilibrium solutions of (\ref{e:wt}) are arbitrary analytic functions \(w = w(z)\) and \(\bm s = 0\). The simplest one is the simple zero, \(w = w_0 =z/\lambda\), the Belavin-Polyakov skyrmion of energy \(E_{xc} =  4\pi J\) and charge \(Q=1\), centered at the origin and of characteristic size \(\lambda\in \mathbb{R}\) (spin up at the origin and spin down at infinity). It is worth noting that if \(w\) is a solution of (\ref{e:wt}), due to the special properties of the stereographic transformation, \(1/w\) is also a solution with \((S_x\rightarrow S_x,S_y\rightarrow -S_y, S_z\rightarrow -S_z)\), and \(1/\bar w\) is a solution with  \((S_x\rightarrow S_x,S_y\rightarrow S_y, S_z\rightarrow -S_z)\), provided that and an identical change is made for the itinerant spins components.

In order to investigate how the spin torque perturbs the skyrmion state \(w_0 = z/ \lambda\), we focus on two simple limiting cases: first, small deviations from the skyrmion state by a uniform polarized current \(\bm s = (0,0,s_z)\); and second, a small circular region around the skyrmion core relevant to track the transition towards the ferromagnetic state \(|w|\rightarrow \infty\). 

In the first case, we linearize (\ref{e:wt}) around the skyrmion state, \(w=z/\lambda+f(z,\bar{z},t)\):
\begin{equation}
  \label{e:ft}
  \I \partial_t f = -J \partial \bar{\partial} f + 
  \frac{2J \bar z}{\lambda^2+|z|^2} \bar \partial f + 
  \frac{s_z}{\lambda} z\,,
\end{equation}
and the spin torque appears as a source term in this approximation. An interesting particular solution is readily found: 
\[
  f=f_0(z,t) = -\I \frac{s_z t}{\lambda}\, z \,.
\]
The pure imaginary factor has the effect of changing the orientation of the magnetization field around the center of the skyrmion, passing successively in time from left to right chirality (as shown in Fig.~\ref{f:skyr}). A sequence observed in the numerical simulations, in addition to the fact that the effective size of the core reduces: \(\lambda\rightarrow\lambda/\sqrt{1+(s_z t)^2}\) (even if, for long times, the perturbation analysis ceases its validity). 

We turn now to the second case, for which we assume that the collapse of the skyrmion core can be described, in the limit of large \(|w|\), by Eq.~(\ref{e:wt}) in the form,
\begin{equation}
  \label{e:wt1}
  \I \partial_t w + J \nabla^2 w = 
  \frac{2J}{w} (\nabla w)^2 + \tfrac{1}{2} s_- w^2
\end{equation}
where, in polar coordinates, 
\[
  \nabla^2 = \frac{1}{r}\partial_r\, r\, \partial_r + 
      \frac{1}{r^2} \partial_{\phi\phi}\,, \quad 
      \nabla = \big(\partial_r , \tfrac{1}{r} \partial_\phi \big)\,.
\]
and the stereographic function is given by \(w=w(r,\phi,t)\), with \(r=|z|\) and \(\phi=\arg z\). We assume that this equation is valid in the neighborhood of the skyrmion core. As the change of topology is, as revealed by the numerical computations, a local processes, studying the behavior of the magnetization in the core region should be enough to exhibit its essential features. In addition, the sudden change in the topology suggests that the main physical mechanism should be the disappearance of the characteristic length scale associated with the skyrmion core size, and therefore, a self-similar evolution in a finite time.

In the asymptotic limit of very large \(|w|\), we search for a solution to (\ref{e:wt1}), appropriated to account for the skyrmion collapse, in a self-similar form:\cite{Zakharov-1972ty}
\begin{equation}
  \label{e:ansatz}
  w(r,\phi,t) = \frac{1}{(t_*-t)^\alpha} 
      f\left(\frac{r}{(t_*-t)^\beta},\phi\right)\,
\end{equation}
which describes the approach of \(w\) to infinity, when \(t \rightarrow t_*\), where \(t_*\) is the collapse time. At \(t=t_*\) the characteristic size of the skyrmion vanishes. Inserting the ansatz (\ref{e:ansatz}) into (\ref{e:wt1}) one obtains two conditions which determine the unknown exponents:
\begin{equation}
  \label{e:ab}
  \alpha=1\,, \quad
  \beta = 1/2\,.
\end{equation}
It is worth noting that the exchange interaction, which is scale invariant, does not permit selecting the \(\alpha\) exponent; its value is fixed by the coupling term with the spin polarized current. A crude estimation of the finite time singularity, which depends in particular on the initial condition, is \(t_* \sim \lambda / s_0 a\), where \(s_0\sim n_e B_p\) is the typical itinerant spin strength per site (which is in the range of \(t_* \sim 10^3\) for our numerical parameters).

The equation of motion satisfied by the self-similar function \(f(X,\phi)\) defined by (\ref{e:ansatz}), where \(X=r/|t_*-t|^{1/2}\), is,
\begin{multline}
  \label{e:f}
  \I \big(1 + \tfrac{1}{2} X \partial_X \big)f = 
      -J \bigg(\partial_{XX} +
      \frac{1}{X} \partial_X + 
      \frac{1}{X^2} \partial_{\phi\phi} \bigg) f \\ + 
      \frac{2J}{f} \bigg[(\partial_X f)^2 + 
      \frac{1}{X^2} (\partial_\phi f)^2 \bigg] + 
      \frac{s_-}{2} f^2\,,
\end{multline}
with the boundary condition \(f(0)=0\) (so \(S_z=1\) at the origin of the skyrmion). Near the origin, the form \(w\sim z\), corresponds to \(f\sim X \E^{\I \phi}\). Near the singularity, the self-similar variable \(X\rightarrow \infty\), the last nonlinear term can be neglected [its main contribution is to the prefactor in (\ref{e:ansatz}), allowing a balance between the time derivative and the nonlinear driving force]. The solution of (\ref{e:f}) can hence be written as 
\begin{equation}
  \label{e:subs0}
  f(X,\phi) = \E^{\I \phi} F(X)\,,
\end{equation}
where \(F(X)\) satisfies the equation,
\begin{multline}
  \label{e:fx}
  \frac{\I}{J} F(X) + \frac{\I}{2J} X \partial_X F(X) + \partial_{XX} F(X)\\
  - \frac{2}{F(X)} [\partial_X F(X)]^2 + 
    \frac{\partial_X F(X)}{X} + \frac{F(X)}{X^2} =0\,.
\end{multline}
The fact that this equation is invariant under scale transformations, suggests the substitution
\begin{equation}
  \label{e:subs1}
  F(X)=C_1 \E^{\,\int \d{X} G(X)}\,,
\end{equation}
with \(C_1\) as an arbitrary complex constant. This substitution enables us to integrate (\ref{e:fx}) once,
\begin{equation}
  \label{e:G}
  \partial_X G(X) + \frac{G(X)}{X} + \frac{1}{X^2} =
       G(X)^2 - \frac{\I X}{2J}G(x) - \frac{\I}{J} \,.
\end{equation}
The solution of (\ref{e:G}) can be found in terms of the Meijer functions, \(G^{m,n}_{p, q}(z; a_1, \dots,a_p; b_1,\dots,b_q)\):\cite{Olver-2010fk,Wolfram-2014kx}
\begin{equation}
  \label{e:solg}
  G(X) = -\frac{\I X}{2 J}
    \frac{ G_{2,3}^{2,1}
    \left(\displaystyle\frac{\I X^2}{4 J}%
      \left|
        \begin{array}{c}
        -1,1 \\
        -\frac{3}{2},-\frac{1}{2},0 \\
        \end{array}%
      \right.
    \right)}
    { G_{1,2}^{2,0}%
      \left(\displaystyle\frac{\I X^2}{4 J}%
        \left|
          \begin{array}{c}
          2 \\
          -\frac{1}{2},\frac{1}{2} \\
          \end{array}%
      \right.
    \right)}\,,
\end{equation}
where an integration constant was chosen to be zero, in order to ensure the correct behavior near \(X=0\). 

Setting (\ref{e:solg}) into (\ref{e:subs1}), and using (\ref{e:ansatz}), and (\ref{e:subs0}), we finally obtain the self-similar solution of the driven Landau-Lifshitz equation in the stereographic projection representation (\ref{e:wt}):
\begin{equation}
  \label{e:solw}
  w(r,\phi,t)= \frac{C_1 \E^{\I \phi}}{|t_*-t|}
    G_{1,2}^{2,0}\left(\frac{\I r^2}{4 J|t_*-t|}\left|
      \begin{array}{c}
        2 \\
        -\frac{1}{2},\frac{1}{2} \\
      \end{array}\right.
    \right)^{-1}
\end{equation}
A fit of (\ref{e:solw}), useful to get some feeling on the structure of this function, can be obtained using the asymptotics of \(G(X)\). These asymptotics can easily be derived from the differential equation (\ref{e:G}) for small and large \(X\). Near the origin \(X\rightarrow 0\), we have \(G(X) \sim 1/X\), and at infinity \(X \rightarrow \infty\), the solution approaches \(G(X) \sim \I X/2J\), leading to the expression,
\begin{equation}
  \label{e:solwa}
  w(r,\phi,t) \approx  \frac{C_1(1+\I) r \E^{\I \phi}}{|t_*-t|^{3/2}}
  \E^{\I r^2/4J|t_*-t|}\,,
\end{equation}
[with a redefinition of the global constant, including a dimensional dependence on \(J\), \((3/8)(\pi/2J)^{1/2} C_1 \rightarrow C_1\)]. This form allows seeing that, at a fixed time, near the origin the solution match a skyrmion \(|w| \sim r/\lambda(t)\), with
\[
  \lambda(t) \sim |C_1|^{-1}|t_*-t|^{3/2}\,.
\] 
Therefore, the size of the skyrmion core vanishes as a power law in time with exponent \(3/2\).  

%
\begin{figure*}
  \centering
  \includegraphics[width=0.33\textwidth]{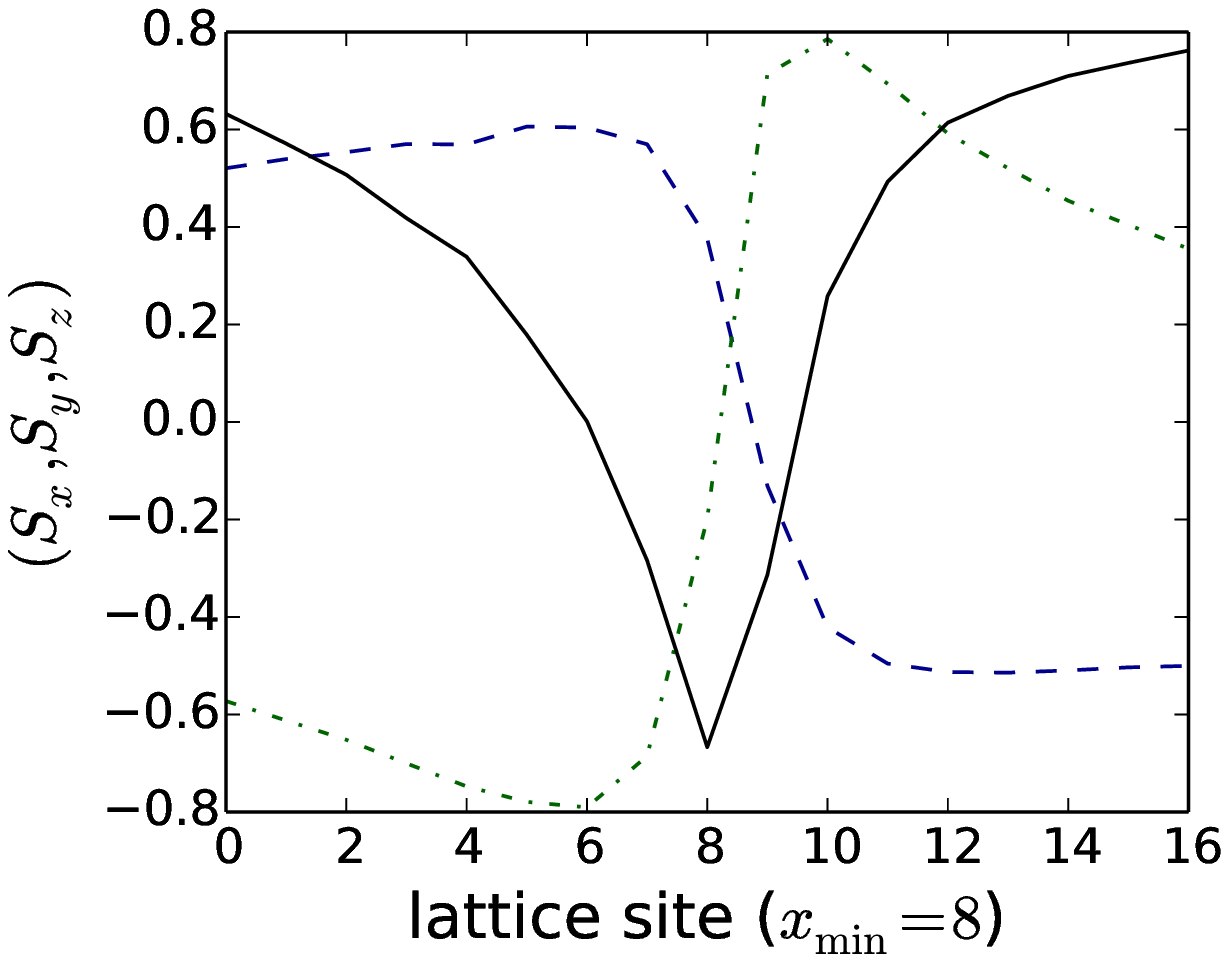}%
  \includegraphics[width=0.38\textwidth]{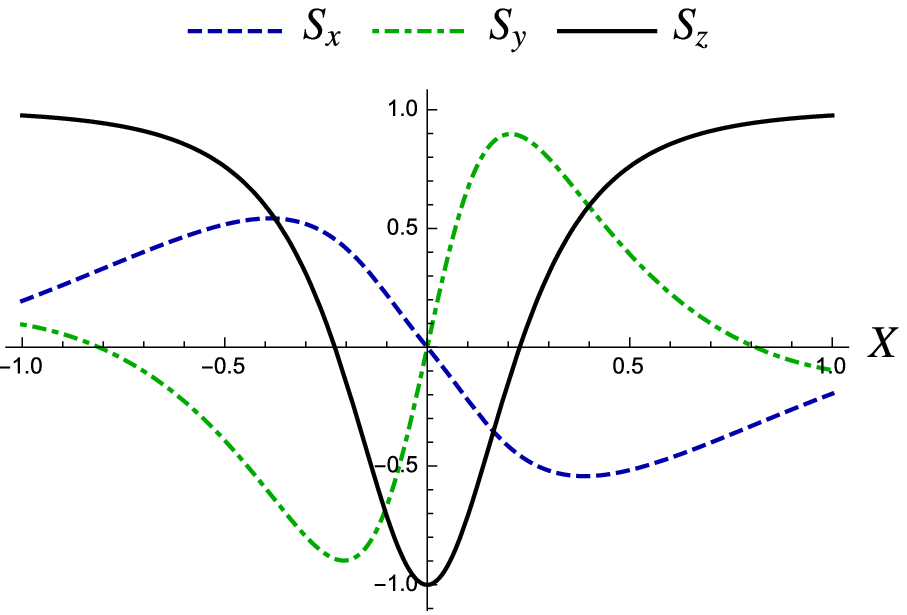}%
  \includegraphics[width=0.27\textwidth]{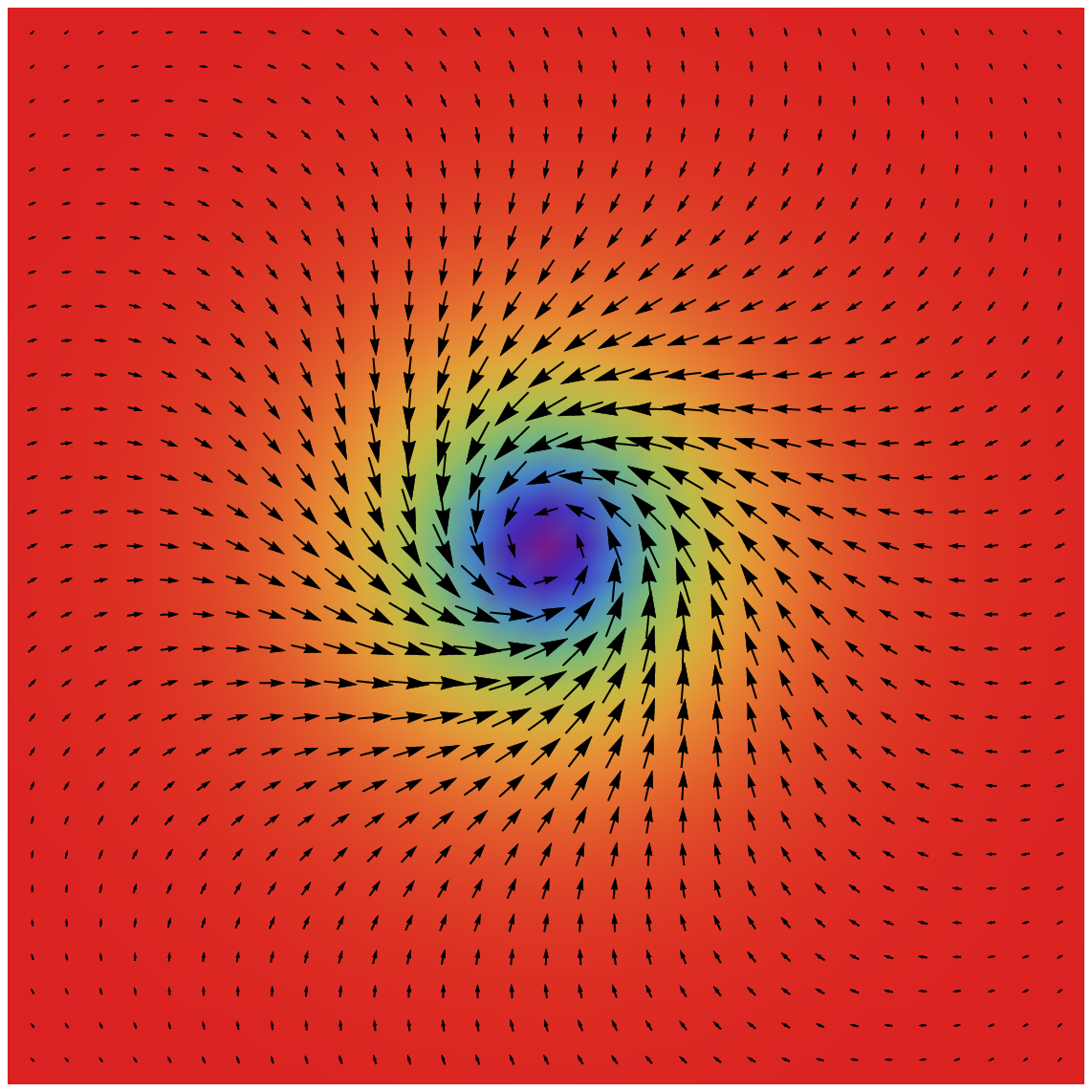}
  \caption{\label{f:ss} Magnetization components corresponding to the numerical computation \(\beta=0.001\), \(t=5936\,t_0\) (left), and to the analytical self-similar solution (center), near the skyrmion core from Eqs.~(\ref{e:solw}) and (\ref{e:istereo}). Analytical magnetization texture (right), to be compared with Fig.~\protect\ref{f:tt}. The size of the core shrinks to zero as the time approaches the critical singularity.}
\end{figure*}

The exact self-similar solution (\ref{e:solw}) is represented in Fig.~\ref{f:ss}, for \(|t_*-t|=0.3\) and \(C_1 =\E^{\I\pi/3}\). Changing the phase of the constant \(C_1\), changes the tilt of the magnetization. The similarity with the numerical computed shape is remarkable. Indeed, for a simple skyrmion form \(w=c z\), with \(c \in \mathbb{C}\) (as the ones shown in Fig.~\ref{f:skyr}), the graph of \(\bm S\) significantly differs with the shape resulting from a position dependent phase. In fact, the form \(\arg w \sim r^2\), allows for distinct amplitudes of the \(S_x\) and \(S_y\) components, breaking the symmetry they possess in the simple skyrmion. In addition, the self-similar form (\ref{e:ansatz}) preserves the stationarity of the exchange energy (\ref{e:excq}) and topological charge (a scale transformation shows that the integrals are independent of time); in particular, the topological charge of the solution (\ref{e:solwa}) is \(Q=1\) (as in the initial state). This is not mandatory for the validity of the calculation, as the explicit solution was deduced assuming a locality condition. The asymptotic evolution depends essentially on the behavior near the singular point (where the exchange interaction is stronger), whereas the topological charge is a global property of the magnetization field. However, it is in accordance with the approximation consisting of keeping only the leading term in a multipole expansion \(\E^{\I m \phi}\), with \(m=1\).

\section{Discussion}

After the experimental evidence of skyrmion phases in chiral magnets,\cite{Muhlbauer-2009vn} phases that can reach the atomic scale,\cite{Heinze-2011fk} it was soon realized that spin-torques are efficient in driven skyrmion dynamics at relatively low current densities.\cite{Jonietz-2010ly} In addition to the spin-transfer torque mechanism, other skyrmion generation mechanisms are possible, for instance, in ferrimagnetic thin films, photo-irradiation leads to a variety of topological configurations that can be controlled by tuning the laser fluence.\cite{Finazzi-2013kx} Discontinuities in the magnetization can also be important.\cite{Iwasaki-2013uq} A domain wall driven by a current can be converted into a skyrmion when entering a wider region,\cite{Zhou-2014kq} an interesting effect in view of its possible application to magnetic memories. A property common to these generation mechanisms of non-trivial magnetic configurations, is the presence of strong inhomogeneities.\cite{Tchoe-2012uq} In our simulations these strong inhomogeneities are spontaneously created by the interaction of the electrons and the magnetization texture. This effect is a consequence of the electron interference and scattering in the potential of fixed spins. In fact, the electron current acts as a random torque on the (classical) magnetic texture. 

\begin{figure}
  \centering
  \includegraphics[width=0.84\linewidth]{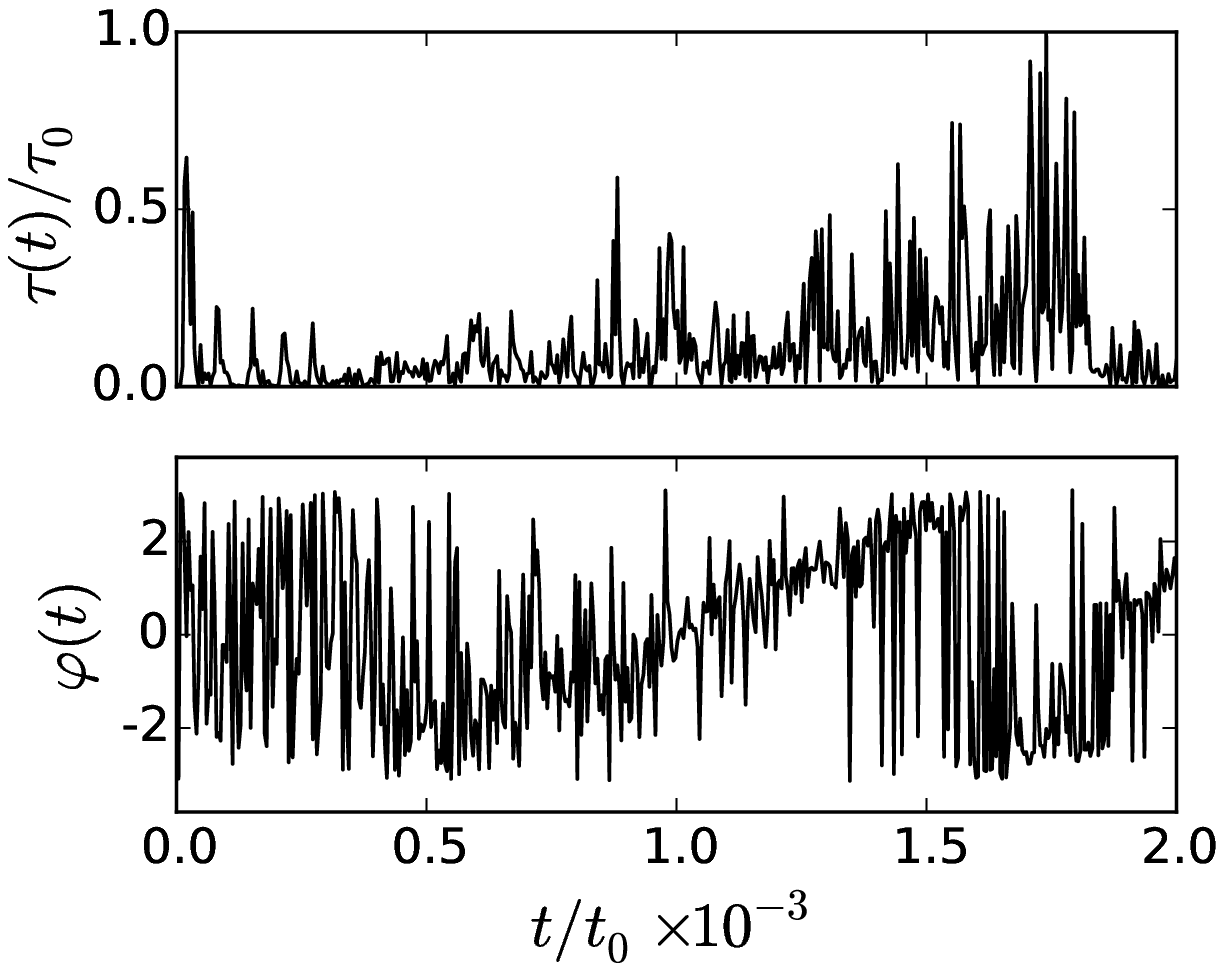}(a)\\
  \includegraphics[width=0.84\linewidth]{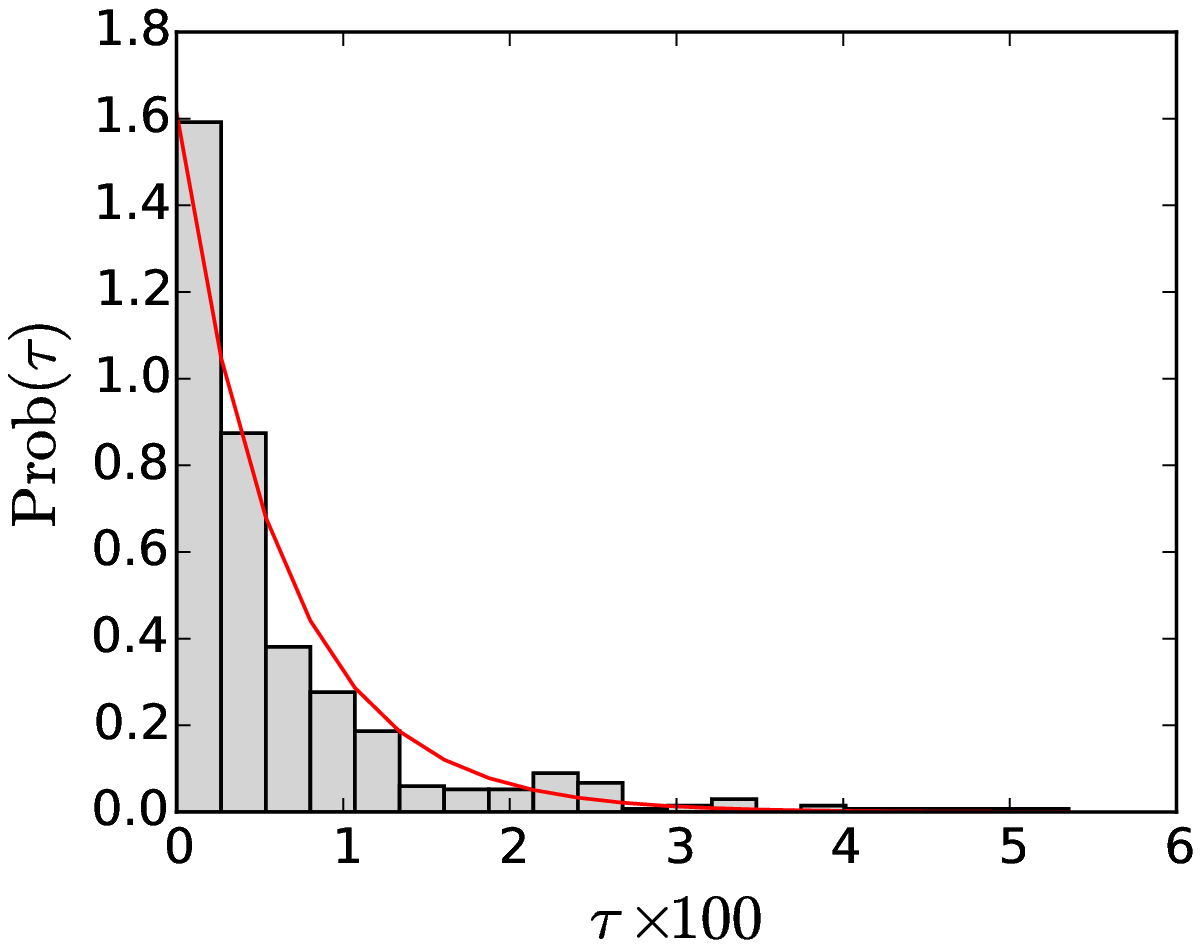}(b)
  \caption{Stochastic torque. The magnitude \(\tau(t)\) and polar angle \(\varphi(t)\) of the electron torque \(\bm \tau(\bm x, t) = \bm s(\bm x,t) \times \bm S(\bm x,t)\) at the skyrmion core \(\bm x = \bm x_0(t)\) as a function of time exhibits a random dynamics (a) with a Poissonian distribution (b), as shown by the exponential fit (red line). Parameters correspond to the intermediate dissipation case, \(\beta=0.01\).
  \label{f:fig6}}
\end{figure}

It might be important to emphasize that the Landau-Lifshitz equation coupled with the Schrödinger equation results in a effective stochastic dynamics of the magnetization. We show in Fig.~\ref{f:fig6} the signal of the spin-transfer torque computed at the skyrmion core position. The highly intermittent dynamics of the torque intensity (defined as its absolute value) corresponds to a Poisson processes, as verified by its histogram. Although other sources of stochasticity can be introduced, for example to take into account thermal effects, it is important to isolate the contribution of the conduction electrons. The relative importance of these mechanisms will depend obviously on the particular experimental situation. In the present case, we observe that the maximum of the torque occurs around the transition time (\(t/t_0\approx 1750\)).

%

During their passage through the skyrmion field, conducting electrons accumulate a Berry phase that leads to interesting electrodynamical effects.\cite{Bruno-2004zr,Schulz-2012yq,Nagaosa-2013qf} The \(b\)-field is a manifestation of such effects. The strong precession of the magnetization in a neighborhood of the skyrmion core, can be interpreted as a reaction to this topological field. The important question here, is whether the \(b\)-field is robust against structural perturbations, notably dissipation. If it is a genuine topological effect, it must be robust under modifications of the dynamics. We accumulated some numerical evidence that under exchange dissipation, the electron vortices are always present during the topological change. 

The dynamics of the skyrmion-ferromagnetic phase transition driven by a polarized current, as we demonstrated in the previous section, is dominated by the self-similar collapse of the skyrmion core. This evolution is universal in the sense that it does not depend on the large scale properties of the system or on the precise values of the physical parameters. At variance to this scenario, the opposite processes, the nucleation of a skyrmion from a current pulse\cite{Tchoe-2012uq} depends on the geometry, characteristic space, and time scales of the driving force. Resulting from a transient, it is not amenable to a similarity solution. Nevertheless, the interaction of the injected electrons with the background magnetization, should nucleate electron vortices that might play a role in the generation of the skyrmion. This problem needs further investigation.

\section{Conclusion}

We investigated the transition between a skyrmion state and a ferromagnetic state driven by a spin polarized electron current. The main goal was to identify the physical mechanism of the topological change and its relation with the dynamics of the skyrmion core. We numerically showed that the torque exerted by the itinerant spins modifies the distribution of the magnetization around the skyrmion core, and tends initially to reduce its size and ultimately drives a topological change. This topological change strongly depends on the dissipation strength. In the absence of dissipation the collapse time explicitly depends on the lattice cutoff. At variance, in the case where dissipation is effective, the transition tends to regularize, and the lifetime of the skyrmion state reduces with increasing dissipation. However, the microscopic mechanism of topological change is in the dissipative case, similar to the nondissipative one (even if the time and length scales may differ). It is related to the appearance of a peculiar electronic structure possessing a net charge opposite to the one in the skyrmion. The synchronization of the fixed spins with this electron vortical structure, leads to the annihilation of the topological charge, which passes from its initial value \(Q=-1\) (skyrmion state) to zero (homogeneous ferromagnetic state).

Using the stereographic projection to represent the Landau-Lifshitz equation in the complex plane,  we could analyze the dynamics of the transition. The polarized current induces a precession of the skyrmion core and tends to reduce its size, as shown by a linear perturbation calculation. When the system is dominated by the exchange interaction and the polarized current, in the strongly nonlinear regime we found that the magnetization follows a self-similar evolution. The self-similar characteristic exponents, which govern the dependence in time of the core size, are determined by the electron spin driving terms. The structure of the self-similar core is essentially determined by the exchange gradients. Applying these approximations, we solved the asymptotic dynamics and obtained a collapse of the core in a finite time. The explicit shape of the self-similar solution compares well with the numerically computed one. In conclusion, this result establishes a link between the topological change and the existence of a finite time singularity in the skyrmion dynamics.

\begin{acknowledgments}
We thank R. G. Elías and S. Rica for useful discussions. 
\end{acknowledgments}

\end{document}